\documentclass[lettersize,journal]{IEEEtran}
\usepackage{amsmath,amsfonts}
\usepackage{algorithmic}
\usepackage{algorithm}
\usepackage{array}
\usepackage[caption=false,font=normalsize,labelfont=sf,textfont=sf]{subfig}
\usepackage{textcomp}
\usepackage{stfloats}
\usepackage{url}
\usepackage{hyperref}
\usepackage{verbatim}

\usepackage{graphicx}
\usepackage[caption=false]{subfig}

\usepackage{svg}
\usepackage{cite}
\usepackage{comment}

\usepackage{soul}
\usepackage{booktabs}
\usepackage{makecell}
\usepackage{colortbl}
\usepackage{multirow}

\usepackage{todonotes}
\setuptodonotes{fancyline, color=blue!30, size=\footnotesize}

\makeatletter
\def\ps@IEEEtitlepagestyle{
    \def\@oddfoot{\customfootnote}%
    \def\@evenfoot{}%
}
\makeatother

\newcommand{\customfootnote}{\footnotesize To appear on IEEE Communications Magazine,  \href{https://doi.org/10.1109/MCOM.002.2400382}{DOI:10.1109/MCOM.002.2400382} \hfill}

\begin{document}

\title{GENIO: Synergizing Edge Computing with\\ Optical Network Infrastructures}

\author{
    \IEEEauthorblockN{Carmine Cesarano\textsuperscript{1}, Alessio Foggia\textsuperscript{1}, Gianluca Roscigno\textsuperscript{2}, Luca Andreani\textsuperscript{3}, Roberto Natella\textsuperscript{1}}
    
    \IEEEauthorblockA{\textit{\textsuperscript{1}Università degli Studi di Napoli Federico II, Naples, Italy}}
    
    \IEEEauthorblockA{\textit{\textsuperscript{2}System Management S.p.A, Naples, Italy}}

    \IEEEauthorblockA{\textit{\textsuperscript{3}DigitalPlatforms S.p.A, Rome, Italy}}
}

\maketitle

\begin{abstract}
Edge computing has emerged as a paradigm to bring low-latency and bandwidth-intensive applications close to end-users. However, edge computing platforms still face challenges related to resource constraints, connectivity, and security. We present GENIO, a novel platform that integrates edge computing within existing Passive Optical Network (PON) infrastructures. GENIO enhances central offices with computational and storage resources, enabling telecom operators to leverage their existing PON networks as a distributed edge computing infrastructure. Through simulations, we show the feasibility of GENIO in supporting real-world edge scenarios, and its better performance compared to a traditional edge computing architecture.
\end{abstract}


\section{Introduction}
Traditionally, edge computing and broadband access networks have been separated from each other. Edge computing platforms focus on running IT applications and services closer to users, either on their devices or on local fog servers. These platforms enable low-latency applications and have been used in several domains such as smart cities, smart infrastructure, and environmental monitoring. On the other hand, access networks provide access for end users to the Internet, ensuring high uplink and downlink speeds. Passive Optical Networks (PONs) have emerged as a standard among telecom operators for access network technology \cite{effenberger2009passive}.

In practice, deploying an edge computing platform still faces significant challenges. Edge devices provide limited computational and storage resources, which can be a bottleneck in computationally intensive applications. Using fog servers is also difficult because they either require dedicated servers and connectivity (e.g., on a local wired or wireless network), which incurs significant costs, or they have to be connected through the Internet, which hampers low latency. 
Furthermore, security and privacy are critical concerns because sensitive data at the edge are exposed to physical and digital attacks. These threats place an additional burden on the security and management of edge and fog servers.

To address these challenges, telecom operators are exploring the placement of applications in central offices (i.e., physical hubs to which users connect) using the telecom infrastructure as an edge computing system \cite{tim_edge_cloud, telefonica_whitepaper}. 
Previous studies have been focused primarily on theoretical design of algorithms for scheduling edge computing tasks (such as, using reinforcement learning) \cite{Ahmad2024, newaz2017}. However, these studies do not consider other aspects that are related to the deployment, orchestration, and management of an edge computing platform.

In line with these trends, we propose the \textit{GENIO} platform (\emph{Edge cloud platform enabled by a new intelligent OLT for GPON networks}). GENIO is a joint industry-academia R\&D project that aims to seamlessly integrate edge computing with PON high-speed broadband networks, overcome existing barriers, and bridge the gap between IT applications and users. 
The main goal of GENIO is to enhance elements of the PON infrastructure, particularly Optical Line Terminals (OLT) and Optical Network Units (ONU), with the capability to run edge services with low latency and guaranteed bandwidth. The GENIO platform will be a cost-effective opportunity for telecom operators to provide computing and storage resources using their existing PON infrastructure. ONUs will be extended with smart embedded systems at the user site, and OLTs will evolve into IT servers with computational and storage capabilities. 

In this work, we present a new software architecture for edge computing on PON infrastructures, which integrates well with off-the-shelf software and hardware of OLT systems. In particular, the architecture leverages recent technology for ``network softwarization'', including VOLTHA/SEBA and ONOS. The GENIO architecture introduces several components for orchestration and management, that are deployed over the cloud, edge, and far edge layers. Moreover, the proposal envisions a workflow that includes both providers of IT applications and end-users that use these IT applications through the PON infrastructure. We also discuss security practices and solutions to harden OLTs against physical and digital attacks. Finally, our work analyzes the feasibility of the GENIO project using off-the-shelf CPUs for use cases of industrial relevance, and compares performance with respect to a traditional edge computing architecture.

\section{Background} 
\label{sec:backgroud}

\begin{figure}[t]
  \includegraphics[width=\linewidth]{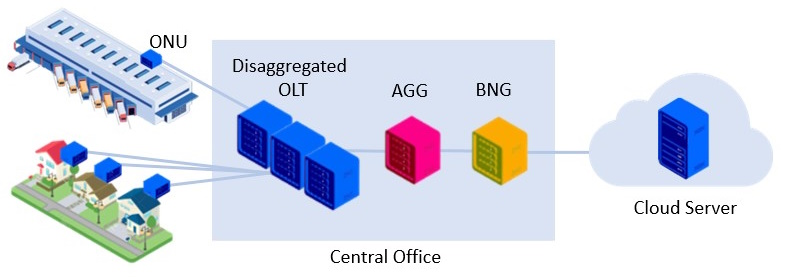}
  \caption{Traditional PON topology for the access network.}
  \label{fig:pon_infrastructure}
\end{figure}

Over the past decade, optical fiber networks have been extensively deployed, particularly in fiber-to-the-home (FTTH) architectures, relying heavily on Passive Optical Network (PON) technologies. Forecasts \cite{broadband_news} suggest that approximately 40 percent of the global population will have fixed FTTH connections by 2026, underscoring the pivotal role of PON technologies. PON networks, with their passive optical distribution networks, have proven to be cost-effective and energy-efficient, with various standardized architectures. Currently, Gigabit-capable Passive Optical Network (GPON) standards, boasting 2.5 Gbps downstream speeds, dominate the landscape. The ongoing evolution towards the XGS-PON standard, promising 10 Gbps in both the upstream and downstream directions, marks the next technological stride. Importantly, these advanced technologies can coexist seamlessly on the same fiber network, ensuring a smooth and efficient transition process. Figure \ref{fig:pon_infrastructure} shows a typical residential PON access network. Users connect through Optical Network Units (ONUs) at their premises, each using dedicated wavelengths for uplink and downlink communications. The PON has a tree structure where ONUs connect via optical fibers to an Optical Line Terminal (OLT) at the Central Office (CO). Within the CO, the Ethernet Aggregation Switch (AGG SW) consolidates traffic from multiple OLTs and directs it to the Broadband Network Gateway (BNG), which connects residential users to the Internet Service Provider (ISP), managing internet access and traffic routing.

In the context of FTTH and PON access networks, communities such as the Open Networking Foundation (ONF) have proposed reference architectures to facilitate the adoption of new PON technologies in a standardized way. For example, SDN-Enabled Broadband Access (SEBA) \cite{das2021cord} represents a paradigm shift in the last-mile solutions for telecom providers. Traditionally, Central Offices have housed proprietary vendor equipment, leading to operational complexities and barriers to innovation. SEBA introduced an innovative architecture that replaces vendor-specific hardware with bare metal components controlled by software with cloud principles and logic. Central Office Re-architected as a Datacenter (CORD) \cite{peterson2016central} is another reference architecture that leverages the principles of Network Function Virtualization (NFV) to reshape the traditional central office, transforming it into a programmable and scalable point of delivery for networking services. Adopting commercial off-the-shelf (COTS) networking equipment and SDN principles in CORD and SEBA architecture empowers operators with centralized configuration, automation, and troubleshooting capabilities, reducing the total cost of ownership.

Virtual Optical Line Termination Hardware Abstraction (VOLTHA) \cite{voltha_doc} serves as a key software component in SEBA and CORD, establishing a unified control and management framework for OLTs and ONUs. VOLTHA introduces an abstraction layer over vendor-specific white-box PON hardware devices. Its northbound interface abstracts the PON access network as a programmable Ethernet switch to an SDN controller based, for example, on the Open Networking Operating System (ONOS) \cite{onos}. On the southbound side, VOLTHA communicates with PON devices using vendor- or device-specific OLT and ONU adapters. In addition, this abstraction layer allows for the separation of OLT functions into distinct hardware and software components, leading to the creation of the so-called disaggregated OLT (dOLT). 

Although existing architectures, such as SEBA and CORD, efficiently manage and control network functionalities, they lack integration with computing capabilities to run applications at the edge of the network. To address this gap, the GENIO platform aims to enhance PON networks with application-management capabilities.

\section{GENIO Platform Architecture}
\label{sec:architecture}

\subsection{Overview}
\begin{figure*}[t]
\centering
\includegraphics[width=\linewidth]{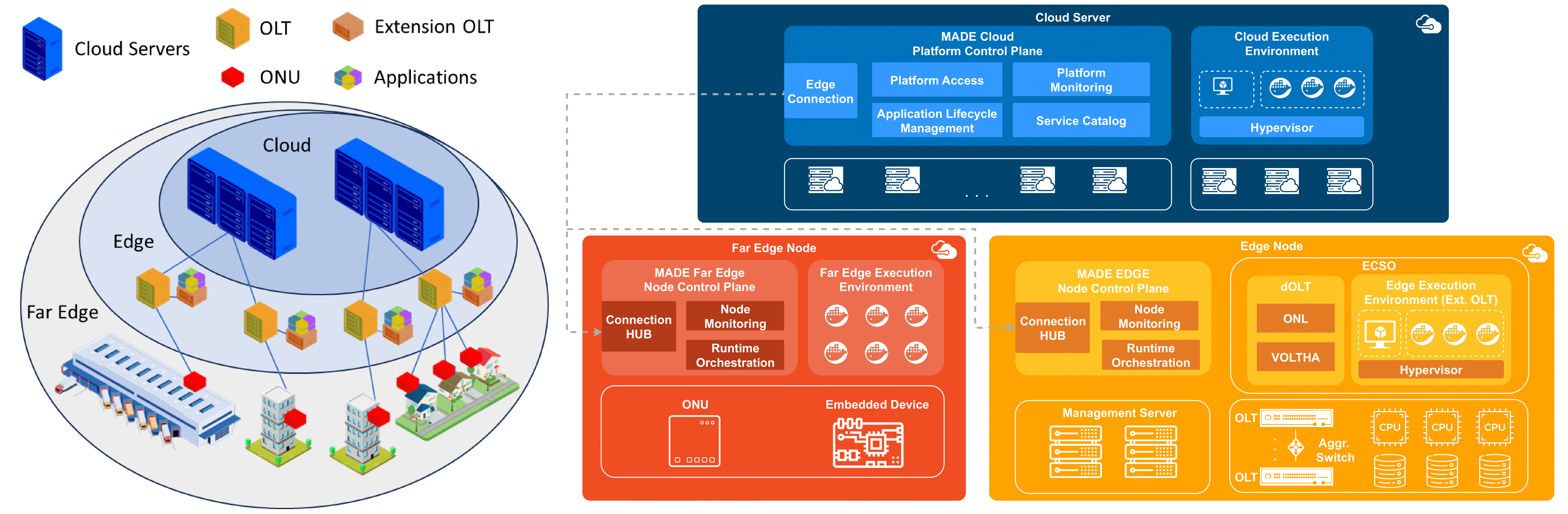}
\caption{On the left, the deployment of GENIO across cloud, edge, and far edge layers. On the right, the GENIO software architecture, which includes the MADE and ECSO components}.
\label{fig:genio_architecture}
\end{figure*}

The GENIO platform is deployed on a distributed architecture across three main layers, as shown in Figure \ref{fig:genio_architecture} (left): the cloud, edge, and far edge. This architecture embodies the principles of edge computing by distributing computational and storage capabilities throughout the network infrastructure, bridging the gap between edge users and the cloud. 

In the far-edge layer, positioned farthest from the cloud server, residential or business users join the platform via ONUs over xGPON network channels. GENIO equips ONUs with low-end computing hardware, extending their capabilities as edge resources for hosting ultralow-latency lightweight applications. Users connect to the applications hosted on GENIO through IoT devices such as smart meters, sensors, and cameras, which can request computing tasks at the edge layer.

The edge layer consists of several dOLTs, which are the endpoints to which the ONUs connect. The dOLT comprises a set of white-box leaf OLT access units connected through an aggregation switch. This scheme enables the grouping of several users at a single Point of Distribution. OLTs are natively designed to provide networking capabilities in PON technology for the telco Central Offices. They are extended for GENIO, by adding off-the-shelf hardware components and new software interfaces to provide computing and storage capabilities.

Finally, each dOLT is connected to the WAN network to provide Internet connectivity, including access to services in the cloud layer. GENIO uses a manager-worker architecture, where the cloud layer provides centralized orchestration of workers deployed at the edge or far-edge layers. In addition, the cloud can deliver computational capabilities for computationally intensive tasks offloaded from other layers.

The design of GENIO includes two fundamental logical subsystems: Multi-Access Edge Cloud (MADE) and Extended Computing node and Secure OLT (ECSO), as shown in Figure \ref{fig:genio_architecture} (right). MADE is the control and orchestration component of the platform and is distributed in the edge and cloud layers. ECSO is a software component that provides worker capabilities to the nodes on the edge under the orchestration of the MADE.

\subsection{MADE Cloud}
MADE Cloud acts as the \textit{platform control plane} and provides access, monitoring, and resource management capabilities for the GENIO platform. Deployed on a centralized cloud server, MADE Cloud provides a global view of all the connected nodes and running applications, orchestrating resources across the entire infrastructure to maintain platform-wide optimizations.

\textbf{Platform Access}. MADE Cloud facilitates multivendor access to the GENIO platform via a standardized interface, supporting two primary business models. First, telecom operators can lease GENIO capabilities to run IT applications for their customers (e.g., vendors of smart infrastructure systems), adopting a provisioning model akin to Infrastructure-as-a-Service (IaaS). This includes provisioning of core resources such as storage, networking, and computing, with telecom operators retaining full control over resource allocation and management, encompassing the operating systems and middleware. Alternatively, business end users can access the platform directly in a software-as-a-service (SaaS) model, utilizing pre-configured applications (e.g., streaming services) with automated scaling and resource adjustments managed by GENIO.

\textbf{Service Catalog}. MADE Cloud maintains a Service Catalog for managing IT applications, which may be customer-supplied or preconfigured. Customers can browse available applications, each described by a Service Profile. This profile serves as a service-level agreement (SLA), specifying the customer's expectations for the hosting environment's quality. The service profile includes parameters such as availability, fault tolerance, response time, bandwidth, memory, number of virtual cores and replicas, isolation and security levels, and energy efficiency.

\textbf{Platform Monitoring}. MADE Cloud implements a two-tier monitoring approach to maintain Quality of Service (QoS). First, it tracks key performance indicators (KPIs) for the infrastructure, covering availability, resource usage (CPU, memory, and storage), network latency, and system reliability (measured by failure rate), and energy consumption. Second, it monitors application-specific KPIs such as response time, throughput, error rate, and concurrent user capacity. Data for both levels were collected from the distributed infrastructure, including both edge and far edge instances. 

\textbf{Application Scheduling}. MADE Cloud autonomously manages applications throughout their lifecycle. This includes both actions for application deployment and actions at runtime to satisfy the service profile requirements. The MADE Cloud will adopt intelligent algorithms for 1) Adaptive Placement, choosing existing or creating new execution environments; 2) Dynamic Scaling, adjusting resource allocation for existing environments; 3) Horizontal Migration, balancing load, and occupancy by migrating across physical nodes within the same layer; and 4) Vertical Migration, moving computationally intensive applications to more capable layers, such as from edge to cloud (i.e., cyberforaging). These actions distinguish GENIO as a QoS-aware and self-adaptive platform, optimizing application performance and ensuring efficient resource utilization in terms of energy, load balancing, and minimizing service fragmentation. 

\subsection{MADE Edge}
MADE Edge acts as the \textit{node control plane}, providing monitoring and orchestration capabilities for the physical nodes where it is deployed, and seamlessly integrates with MADE Cloud management. Replicas of the MADE Edge can be instantiated on both edge and far edge nodes, enabling the use of OLTs and ONUs as worker nodes for executing edge applications.

\textbf{Runtime Orchestration}. MADE Edge continuously controls and orchestrates deployed applications at runtime using technologies such as Kubernetes. The orchestration process considers offloading capabilities and workload distribution across various containerized computational environments. It incorporates application policies generated from the Service Profile and runtime orchestration, based on real-time data gathered through low-level monitoring. Dynamically adapting to changing node and application conditions, MADE Edge ensures QoS at the node level. 

\textbf{Node Monitoring}. MADE Edge is a crucial component for monitoring worker nodes. It continuously gathers data from both the physical node and applications and transmits them to the centralized MADE Cloud. This low-level monitoring provides the observability of the system and insights into resource utilization, network performance, and system reliability.

\textbf{Resource Management}. MADE Edge manages the computational and networking resources of an Extended dOLT, by interacting with the ECSO component. Moreover, MADGE Edge leverages VOLTHA and ONOS to establish connectivity between edge applications running in the dOLT and end users at the other end of the PON network.

\subsection{Made Far Edge}
MADE Far Edge follows the same MADE Edge architecture, including similar components for runtime orchestration and monitoring. It consists of an ONU extended with computing hardware at the user premises. The MADE Far Edge provides network connectivity to end-users, as in the case of the ONU in traditional PON networks. Additionally, it can run resource-constrained edge applications. Users connect to MADE Far Edge through their IoT devices using a local area network (e.g., Ethernet or WiFi), to send processing requests to edge applications hosted by GENIO. In specific scenarios (e.g., deployments in wide areas), IoT devices can connect to the MADE Far Edge through cellular connectivity, which provides wide coverage and high availability at the cost of higher complexity and latency. 
The MADE Far Edge is designed for energy efficiency, to suit both residential environments, where reduced energy costs are valued, and open environments, where battery-powered devices are often used.

\subsection{Interactions between MADE Cloud and MADE Edge}
MADE Cloud and MADE Edge work together to maintain an adaptive, QoS-driven environment, optimize resources, and ensure fault resilience. 

\textbf{Application Placement.} Upon receiving a placement request, MADE Cloud queries MADE Edge instances for real-time metrics on resource usage, node occupancy, energy consumption, and network performance. Based on load and SLA needs, MADE Cloud selects the optimal deployment node and sends deployment instructions with specified resources and configurations.

\textbf{Resource Scaling and Workload Migration.} MADE Edge monitors local resource demands and alerts MADE Cloud during load spikes. MADE Cloud then adjusts resource allocations within the edge layer or, if needed, shifts workloads to the cloud. 

\textbf{Fault Detection and Recovery.} MADE Edge monitors node health and security, reporting issues to MADE Cloud. MADE Cloud responds by managing failovers, redistributing workloads, or offloading tasks to the cloud layer, ensuring service continuity and enhancing GENIO with proactive fault management.

\subsection{ECSO}
The \textit{Extended Computing node and Secure OLT} (ECSO) is deployed on edge nodes located in central offices. It consists of \textit{dOLT}, which manages PON networking, and \textit{Extension OLT}, which handles processing tasks.

\textbf{dOLT} is the networking component of ECSO, following the \textit{disaggregated OLT} paradigm as discussed in Section \ref{sec:backgroud}. Unlike traditional OLTs, where one vendor bundles all functions (e.g., control, switching, and user interfaces) into a single package, ECSO uses virtualization and software-defined networking (SDN) technology, such as ONOS \cite{onos} and ONL \cite{onl}, to break down OLT functions into separate software modules. Under this approach, various OLTs, whether branded, white-box, or virtual, can coexist in the same central office. They were configured and managed through a unified northbound software abstraction provided by VOLTHA. This flexibility allows network operators to choose different suppliers for components, enhances scalability, and simplifies updates.

\textbf{Extension OLT} enhances the capabilities of dOLT by incorporating computing and storage capabilities using off-the-shelf hardware (e.g., x86 servers), for the execution of third-party applications in the central office. In line with GENIO's vision of multitenancy, applications are executed within isolated runtime environments. Extension OLT offers varying levels of isolation to provide the best trade-off between performance and security according to diverse tenant needs. The provisioning strategy includes both \textit{Soft Multi-Tenancy}, which allows shared resources among trusted users (e.g., using container-based virtualization), and \textit{Hard Multi-Tenancy}, which ensures strict isolation among different tenants (e.g., using virtual machines). This approach enables multiple service providers to share the multitenant GENIO infrastructure.

\section{Security-by-design}
\label{sec:security}
The GENIO project adopts a security-by-design approach to align with the security requirements imposed by international regulations. In particular, the European Cyber Resilience Act (CRA) mandates that ``products with digital elements'' comply with security requirements to be offered on the EU market and to receive the ``CE mark''.

To secure the PON network infrastructure, the GENIO project will conduct a comprehensive threat modeling based on the well-known Microsoft STRIDE approach. The thread model will encompass threats against the PON network, including fiber links, ONUs, and OLTs, which are deployed in the field and are targets of physical attacks, e.g., to tamper with the OLT systems. Moreover, the analysis will encompass threats that arise from multi-tenancy, since edge applications may be compromised by attackers and be used as entry points to attack other applications of other tenants and the GENIO platform itself. Finally, threat modeling will identify threats that arise from the cloud and web.

Threats will be systematically addressed based on the threat model. To address physical threats, the OLTs will be hardened with secure boot technology to assure software integrity. Moreover, the GENIO platform will adopt automatic updates of the OLT firmware, using digital signatures to ensure the integrity of the updates. Digital threats against the OLTs will be mitigated by hardening the OLT firmware. In particular, third-party and open-source components (including libraries, kernels, and orchestrators) will be protected by debloating and hardening their configuration (e.g., disabling unnecessary features, and enabling optional security controls), and by maintaining a Software Bill of Materials for transparency and rapid response to any new vulnerabilities that may be found in these components.

To ensure isolation among tenants, the GENIO platform will provide both soft isolation, by using separate network namespaces and containers to host multiple workloads; and hard isolation, by using full virtualization and separated instances of orchestration components to provide stronger isolation guarantees that users may require. As previously discussed, the hypervisor will be protected from container- and VM-escape attacks by hardening its configuration. Kubernetes, ONOS, and other orchestration components will be secured against Denial of Service and privilege escalation through stringent access controls and resource quotas.

The software components of the GENIO platforms, including MADE Cloud, MADE Edge, and MADE Far Edge, will be developed according to best industry practices software engineering practices, by adopting a secure software development process and systematic quality assurance activities. These components will perform access control to protect resources from unauthorized access, such as the Service Catalog, and will adopt secure communication based on SSL/TLS to prevent network attacks such as sniffing and spoofing. Finally, to mitigate vulnerabilities in edge applications, these components will perform automatic security scans of container images before including the images in the Service Catalog.

\section{Use Cases and Simulation}
\label{sec:simulation_analysis}
We conducted an experimental evaluation with two primary objectives. First, we simulated the performance of GENIO architecture across various commercial off-the-shelf CPUs, to demonstrate the feasibility of common edge computing scenarios. This allows telecom operators to enhance their OLT setups cost-effectively with minimal hardware investment. Second, we compared GENIO with a traditional edge computing architecture, to highlight its performance gain.

\subsection{Edge Computing Scenarios}
\label{sec:scenarios}
In our analysis, we considered four common edge computing use cases including smart cities, e-health, smart buildings, and AI-Generated Content (AIGC) that are of interest to the industrial partners of the GENIO project. 

In smart-city applications, edge computing can be used to improve the efficiency of urban services and promote environmental sustainability through the collection and analysis of sensor data. E-health applications take advantage of edge computing to improve access to and management of healthcare services, improve quality of care, and promote overall well-being. Smart building applications aim to optimize energy efficiency and enhance security by monitoring and managing energy consumption, lighting, and infrastructure access. Finally, AIGC services leverage edge servers to generate image and video content, based on local AI models at the edge and data collected from the end-users.

\begin{figure}[t]
  \includegraphics[width=\linewidth]{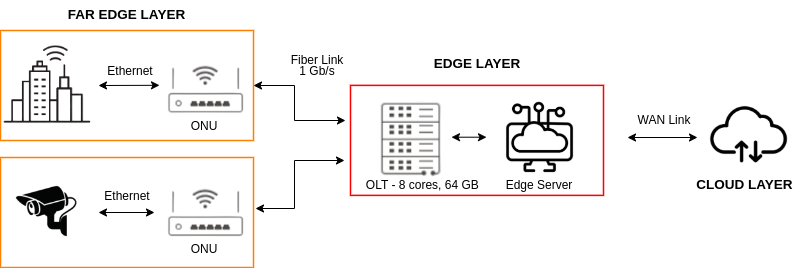}
  \caption{Simulation environment for the GENIO setup in PureEdgeSim}
  \label{fig:genio_simulation}
\end{figure}

\subsection{Simulation environment}
We used the widely adopted PureEdgeSim \cite{mechalikh2019pureedgesim} simulator for our evaluations due to its flexibility in configuring simulated cloud and edge environments. PureEdgeSim enables users to define custom topologies, set node distances, and customize computational nodes, network links, workloads, and orchestration schemes. To simulate and evaluate the GENIO architecture, we extended PureEdgeSim with custom components, including different link types (optical fiber, MAN, and WAN) to model their different latency and bandwidth characteristics. Our customized version is publicly available\footnote{\url{https://github.com/dessertlab/PureEdgeSim}}.

In addition, PureEdgeSim can generate workloads as task groups executed on the simulated edge infrastructure. By adjusting workload parameters, different edge applications can be simulated. These parameters include maximum desired latency, task length measured in Millions of Instructions (MI), task generation rate, request size, and result size. Table \ref{tab:config} summarizes the parameter values for the four simulated edge scenarios. These parameters are based on previous studies that simulated real-world edge applications \cite{collab,mechalikh2019pureedgesim, enhancing2024}, and on information from industrial partners of the project. 
The workloads are configured as follows:

\begin{table}[b]
\caption{Simulation parameters for the application scenarios}
\label{tab:config}
\centering
\begin{tabular}{lllll}
  \toprule
  \textbf{Application} & \textbf{Smart} & \textbf{E-Health} & \textbf{Smart} & \textbf{AIGC}\\
  & \textbf{City} &  & \textbf{Building} & \\
  \midrule
  Users & 128 & 10 & 20 & 50 \\
  Rate (task/min) & 2 & 60 & 60 & 20 \\
  Latency (s) & 0.5 & 0.05 & 0.2 & 0.5 \\
  Task length (MI) & 500 & 1,000  & 5,000  & 48,000 \\
  Request size (KB) & 1 & 10 & 750 & 5000 \\
  Result size (KB) & 10 & 10 & 500 & 2000 \\
  \bottomrule
\end{tabular}
\end{table}

\begin{itemize}
    \item \textbf{Smart City}: 128 smart lights connected through the same OLT, which transmit a 1kB message at a rate of 2 requests per minute. This scenario has a low computational load and can afford a high service latency compared to the other scenarios.
    \item \textbf{E-Health}: 10 cardio monitors transmitting vital parameters in 10kB messages at a rate of 60 requests per minute. It is a latency-sensitive application, with a relatively low computational load. 
    \item \textbf{Smart Building}: 20 video surveillance cameras transmitting post-processed video frames that generate 60 requests per minute, with each request being 750kB in size. It is a computing-intensive application, with moderate latency requirements.
    \item \textbf{AI Generated Content}: 50 users sending requests for generating image contents using an AI model, at a rate of 20 requests per minute, with each request being 5000kB in size. This is the most computing-intensive scenario.
\end{itemize}

\textbf{GENIO setup.} The GENIO simulation environment consists of three layers, as shown in Figure \ref{fig:genio_simulation}. 
The end-user devices at the \emph{far edge layer} generate processing requests for edge applications. The number of devices varies by scenario, as specified in Table \ref{tab:config}. Each device is modeled after a Raspberry Pi 3 Model B+ board and pairs with ONUs connected via zero-distance to simulate a device at the customer premise.
The edge and the far edge layers are connected through direct optical connections, which are modeled as a 1 Gbps link of 100 meters. The edge layer is modeled as an edge server connected via a zero-distance link to the OLT, in order to simulate the deployment of the edge applications in the PON central office. 
The cloud layer, represented by a single cloud server, manages task scheduling. Task scheduling is managed by the TRADE\_OFF algorithm in PureEdgeSim. This algorithm uses information about throughput and task length to effectively distribute tasks across the nodes, optimizing CPU occupation and measured latency. The cloud server connects to the edge server via WAN links, modeled as links of 50 kilometers.

\textbf{Baseline setup.} As a reference for evaluating the GENIO architecture, we configured an additional setup based on a traditional edge computing architecture. It represents the typical topology of PON-based access networks, where the OLT is located at the central office, while the edge server and edge applications are located on a remote server. The edge servers are reached by the OLT through WAN links, 100 meters in length. Therefore, in this setup, the simulation is configured to divide the edge layer into two different layers.

\begin{figure*}[t]
\centering
\includegraphics[width=\linewidth]{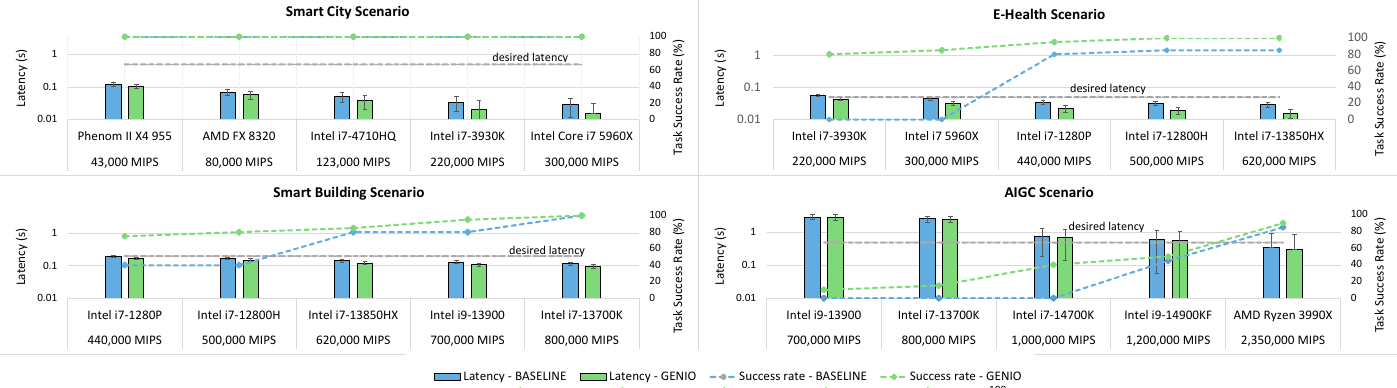}
\caption{Latency and task success rate comparison across four edge scenarios, contrasting GENIO and baseline architectures for different CPU types.}
\label{fig:simulation_results}
\end{figure*}

\begin{table}[b]
\centering
\caption{Selected CPUs for the simulated applications}
\begin{tabular}{lllc}
\toprule
\textbf{Processor} & \textbf{MIPS} & \textbf{Application} \\
\midrule
AMD Phenom II X4 955 & 43,000 & \cellcolor{green!10} Smart City & \\
AMD FX 8320 & 80,000 & \cellcolor{green!10} Smart City & \\
Intel Core i7-4710HQ & 123,000 & \cellcolor{green!10} Smart City & \\
Intel Core i7-3930K & 220,000 & \cellcolor{green!10} Smart City & \cellcolor{blue!10} E-Health \\
Intel Core i7 5960X & 300,000 & \cellcolor{green!10} Smart City & \cellcolor{blue!10} E-Health \\
Intel i7-1280P & 440,000 & \cellcolor{red!10} Smart Building  & \cellcolor{blue!10} E-Health \\
Intel i7-12800H & 500,000 & \cellcolor{red!10} Smart Building & \cellcolor{blue!10} E-Health  \\
Intel i7-13850HX & 620,000 & \cellcolor{red!10} Smart Building & \cellcolor{blue!10} E-Health  \\
Intel i9-13900 & 700,000 & \cellcolor{red!10} Smart Building & \cellcolor{orange!10} AIGC \\
Intel i7-13700K & 800,000 & \cellcolor{red!10} Smart Building & \cellcolor{orange!10} AIGC \\
Intel i7-14700K & 1,000,000 & & \cellcolor{orange!10} AIGC \\
Intel i9-14900KF & 1,200,000 & & \cellcolor{orange!10} AIGC \\
AMD Ryzen 3990X & 2,350,000 & & \cellcolor{orange!10} AIGC \\
\bottomrule
\end{tabular}
\label{tab:processors}
\end{table}

\subsection{Hardware configurations}
We ran simulations with varying hardware configurations for the edge server in both GENIO and baseline setups, analyzing their impact on the performance of edge computing applications. According to our industrial partners, the extended OLT product will incorporate COTS x86 technology to manage both PON networking and edge computing. In the simulations, we varied computing capabilities by setting different MIPS values for the extended OLT. We performed an additional analysis where we found that a minimum of 100 MB storage and 1 MB RAM ensures no performance degradation across all scenarios, as the tasks do not stress storage capacity and are primarily CPU-intensive. Therefore, we fixed the RAM and storage at 32 GB and 1 TB, respectively, since this hardware configuration does not impact performance and is affordable to equip on the extended OLT.

We simulated MIPS values that align with the existing COTS x86 processors from Intel and AMD vendors, spanning entry-level, mid-range, and high-performance models. We adopted the Sandra Drystone benchmark \cite{sandra_drystone} to obtain MIPS values for the chosen commercial processors. For each application scenario, we selected five CPU models. Table \ref{tab:processors} summarizes the selected CPUs and edge applications intended for execution.

\subsection{Experimental Results}
In our experimental analysis, we measured \emph{task latency} and \emph{task success rate} (TSR) for each edge application scenario, over different CPU architectures. Task latency represents the delay between sending a request and its processing; the TSR indicates the percentage of tasks that are completed correctly within the simulation time. These metrics reflect the quality of service achieved by edge computing systems. For each of the scenarios described in Sec. \ref{sec:scenarios}, we expected a 100\% success rate and a \emph{desired latency} up to 0.5, 0.05, 0.2 and 0.5 seconds for Smart City, E-Health, Smart Building and AIGC, respectively (see also Table \ref{tab:config}). Figure~\ref{fig:simulation_results} shows the metrics for each scenario. The bars represent the TSR, while the dots indicate the associated latency across five different CPUs. We evaluated the latency and TSR of both the GENIO setup and the baseline setup. Additionally, the gray dashed lines show the maximum desired latency requirement for each scenario, providing a visual reference to assess performance.
In our experimental analysis, GENIO consistently outperformed the baseline architecture in terms of latency across all CPUs and edge scenarios. For every CPU architecture, GENIO showed a lower latency compared to the baseline, indicating a faster processing time. Specifically, average latency reductions of $-27.7\%$, $-35.0\%$, $-13.1\%$, and $-7.1\%$ were observed for the Smart City, E-Health, Smart Building, and AIGC scenarios, respectively. Furthermore, the TSR in all scenarios was improved under GENIO, as shown by TSR curves that consistently outperformed the baseline, indicating a higher rate of completed tasks. In the case of AIGC services, only the high-end CPUs were able to achieve high TSR and low latency requirements, due to the high computational cost of this type of workload. This scenario would benefit from further computational resources, by adding GPUs to the OLT.

These findings underscore the performance advantage of GENIO, with reduced latency and increased TSR compared to traditional edge servers over the Internet. These results suggest that network operators could feasibly integrate GENIO into existing PON infrastructures using COTS hardware, enabling robust support for a diverse set of edge scenarios.

\section{Conclusion}
\label{sec:conclusion}
This paper presented the GENIO platform, which seamlessly integrates edge computing capabilities within Passive Optical Network (PON) infrastructures, allowing for effectively transforming telecom central offices in a distributed edge computing platform. 
Through the PureEdgeSim simulation framework, we demonstrated the feasibility of GENIO in supporting real-world edge computing scenarios like smart cities, e-health, and smart buildings. Our results show that by augmenting OLTs with modest off-the-shelf hardware, telecom operators can execute latency-sensitive and compute-intensive edge applications with high success rates and low latencies, compared to a traditional edge computing architecture with PON-based access. The design and concepts introduced with GENIO can also be easily extended to edge computing over 5G-based networks, as this change affects only the access layer. This adaptability positions GENIO for seamless integration in next-generation telecom infrastructures.

\section*{Acknowledgments}
This project has been partially supported by the GENIO project (CUP B69J23005770005) funded by MIMIT, "Accordi per l'Innovazione" program.

\IEEEtriggeratref{27}
\bibliographystyle{IEEEtran}
\bibliography{bibliography}

\begin{thebibliography}{10}
\providecommand{\url}[1]{#1}
\csname url@samestyle\endcsname
\providecommand{\newblock}{\relax}
\providecommand{\bibinfo}[2]{#2}
\providecommand{\BIBentrySTDinterwordspacing}{\spaceskip=0pt\relax}
\providecommand{\BIBentryALTinterwordstretchfactor}{4}
\providecommand{\BIBentryALTinterwordspacing}{\spaceskip=\fontdimen2\font plus
\BIBentryALTinterwordstretchfactor\fontdimen3\font minus
  \fontdimen4\font\relax}
\providecommand{\BIBforeignlanguage}[2]{{%
\expandafter\ifx\csname l@#1\endcsname\relax
\typeout{** WARNING: IEEEtran.bst: No hyphenation pattern has been}%
\typeout{** loaded for the language `#1'. Using the pattern for}%
\typeout{** the default language instead.}%
\else
\language=\csname l@#1\endcsname
\fi
#2}}
\providecommand{\BIBdecl}{\relax}
\BIBdecl

\bibitem{effenberger2009passive}
F.~Effenberger and T.~S. El-Bawab, ``{Passive optical networks (PONs): past,
  present, and future},'' \emph{Optical Switching and Networking}, vol.~6,
  no.~3, pp. 143--150, 2009.

\bibitem{tim_edge_cloud}
\BIBentryALTinterwordspacing
{TIM}, ``{Edge Cloud Computing},'' 2022, accessed on 05.02.2024. [Online].
  Available:
  \url{https://www.gruppotim.it/it/newsroom/notiziario-tecnico-tim/2022/n1-2022/cap03-edge-cloud-computing.html}
\BIBentrySTDinterwordspacing

\bibitem{telefonica_whitepaper}
\BIBentryALTinterwordspacing
{Telefonica}, ``{Telefónica Open Access and Edge Computing},'' 2019, accessed
  on 05.02.2024. [Online]. Available:
  \url{https://www.telefonica.com/es/wp-content/uploads/sites/4/2021/02/whitepaper-telefonica-opa-mec-feb-2019.pdf}
\BIBentrySTDinterwordspacing

\bibitem{Ahmad2024}
F.~H. Ahmad, S.~H. Shah~Newaz, D.~S. Sankar, R.~Ramlie, and N.~S. Nafi, ``Fog
  computing in optical access networks: An energy-efficient and deadline-aware
  task scheduling mechanism,'' in \emph{2023 13th International Conference on
  Information Technology in Asia (CITA)}, 2023, pp. 66--69.

\bibitem{newaz2017}
S.~H.~S. Newaz, W.~Susanty~binti Haji~Suhaili, G.~M. Lee, M.~R. Uddin, A.~F.~Y.
  Mohammed, and J.~K. Choi, ``{Towards realizing the importance of placing fog
  computing facilities at the central office of a PON},'' in \emph{2017 19th
  International Conference on Advanced Communication Technology (ICACT)}, 2017,
  pp. 152--157.

\bibitem{broadband_news}
\BIBentryALTinterwordspacing
{OMIDIA}, ``{Fiber and Copper Access Equipment Forecast: 2021-27},'' 2022,
  accessed on 04.02.2024. [Online]. Available:
  \url{https://omdia.tech.informa.com/om022089/fiber-and-copper-access-equipment-forecast-202127}
\BIBentrySTDinterwordspacing

\bibitem{das2021cord}
S.~Das, ``{From CORD to SDN enabled broadband access (SEBA)},'' \emph{Journal
  of Optical Communications and Networking}, vol.~13, no.~1, pp. A88--A99,
  2021.

\bibitem{peterson2016central}
L.~Peterson, A.~Al-Shabibi, T.~Anshutz, S.~Baker, A.~Bavier, S.~Das, J.~Hart,
  G.~Palukar, and W.~Snow, ``Central office re-architected as a data center,''
  \emph{IEEE Communications Magazine}, vol.~54, no.~10, pp. 96--101, 2016.

\bibitem{voltha_doc}
\BIBentryALTinterwordspacing
{ONF}, ``{VOLTHA Documentation},'' accessed on 15.02.2024. [Online]. Available:
  \url{https://docs.voltha.org/}
\BIBentrySTDinterwordspacing

\bibitem{onos}
\BIBentryALTinterwordspacing
{Open Networking Foundation}, ``{ONOS},'' 2020, accessed on 10.06.2024.
  [Online]. Available: \url{https://opennetworking.org/onos/}
\BIBentrySTDinterwordspacing

\bibitem{onl}
\BIBentryALTinterwordspacing
{Open Networking Linux}, ``{ONL},'' 2020, accessed on 25.06.2024. [Online].
  Available: \url{https://opennetworking.org/onos/}
\BIBentrySTDinterwordspacing

\bibitem{mechalikh2019pureedgesim}
C.~Mechalikh, H.~Taktak, and F.~Moussa, ``{PureEdgeSim: A simulation toolkit
  for performance evaluation of cloud, fog, and pure edge computing
  environments},'' in \emph{2019 international conference on high performance
  computing \& simulation (HPCS)}.\hskip 1em plus 0.5em minus 0.4em\relax IEEE,
  2019, pp. 700--707.

\bibitem{collab}
W.~Fan, L.~Zhao, X.~Liu, Y.~Su, S.~Li, F.~Wu, and Y.~Liu, ``{Collaborative
  Service Placement, Task Scheduling, and Resource Allocation for Task
  Offloading With Edge-Cloud Cooperation},'' \emph{IEEE Transactions on Mobile
  Computing}, vol.~23, no.~1, pp. 238--256, 2024.

\bibitem{enhancing2024}
C.~Xu, J.~Guo, J.~Zeng, S.~Meng, X.~Chu, J.~Cao, and T.~Wang, ``{Enhancing
  AI-Generated Content Efficiency Through Adaptive Multi-Edge Collaboration},''
  in \emph{2024 IEEE 44th International Conference on Distributed Computing
  Systems (ICDCS)}, 2024, pp. 960--970.

\bibitem{sandra_drystone}
\BIBentryALTinterwordspacing
{SiSoftware}, ``{Sandra Lite},'' 2020, accessed on 10.06.2024. [Online].
  Available: \url{https://sisoftware-sandra-download.com/}
\BIBentrySTDinterwordspacing

\end{thebibliography}

\section*{Biographies}
\vspace{-35pt}

\begin{IEEEbiographynophoto}{Carmine Cesarano} (carmine.cesarano2@unina.it) is a PhD candidate at the Federico II University of Naples. His research interests are in the Open Source Supply Chain Security.
\end{IEEEbiographynophoto} 

\vspace{-35pt}

\begin{IEEEbiographynophoto}{Alessio Foggia} graduated in Computer Engineering at the Federico II University of Naples, where he currently holds a position as a research assistant. His research interests are in the field of Edge Computing security.
\end{IEEEbiographynophoto}

\vspace{-35pt}

\begin{IEEEbiographynophoto}{Gianluca Roscigno}, PhD, has been PhD graduate and Research Fellow in Computer Science at the University of Salerno. Since 2017, he has been working in the R\&D area at System Management S.p.A. (part of the DigitalPlatforms S.p.A. group), where he currently holds the role of R\&D Funded Projects Manager.
\end{IEEEbiographynophoto}

\vspace{-35pt}

\begin{IEEEbiographynophoto}{Luca Andreani} is an R\&D Manager on Telecommunication Technology at the Innovation Business Unit of DigitalPlatforms SpA. He has over twenty years of experience in design, development, testing, engineering and sales of highly technological solutions for primary national and international customers.
\end{IEEEbiographynophoto}

\vspace{-35pt}

\begin{IEEEbiographynophoto}{Roberto Natella}, PhD, is Associate Professor at the Federico II University of Naples. His research interests are in the field of software security and dependability. In 2022, he received the DSN 2022 Rising Star in Dependability Award from the IEEE TCFT and IFIP WG 10.4.
\end{IEEEbiographynophoto}

\end{document}